\documentclass[11pt]{article}
\usepackage{amsmath}
\usepackage{amssymb}
\usepackage{graphicx}
\usepackage{cite}

\title{Single-Shot Fidelity Reveals Hard and Soft Limits: A Universal Yardstick for Photon-Number-Resolving Detectors}
\author{Tetsuya Tsuruta, Akio Yoshizawa, Daiji Fukuda \\[2pt] \small National Institute of Advanced Industrial Science and Technology (AIST)}
\date{}

\begin{document}
\maketitle

\begin{abstract}
\begin{sloppypar}
Photon-number-resolving (PNR) detectors are central to both discrete-variable (DV) and continuous-variable (CV) photonic quantum computing, where a single measurement outcome must herald a specific Fock state in real time. Yet the fidelity conventionally reported for these detectors comes from a curve fit accumulated over many shots -- a population estimate that converges to the correct answer however badly neighboring photon-number peaks overlap, so a detector can be certified as accurate even though any single trial may be more likely wrong than right. We close this gap by composing a resolution-driven confusion matrix, built directly from a detector's continuous energy response, with its conventional (efficiency-only) POVM, keeping the two origins of error structurally separate while recombining them into a single-shot fidelity that reports, directly, the probability of correctly heralding a given photon number. This separation exposes a stark asymmetry: detection-efficiency loss is an unrecoverable, hardware-fixed floor, whereas resolution-driven misidentification can be converted into a heralded erasure by narrowing the accept region, trading generation rate for confidence. Because the resulting fidelity is defined identically regardless of how a detector encodes photon number, it also puts energy-resolving detectors such as transition-edge sensors -- whose efficiency and resolution contributions have not previously been disentangled -- on equal footing with multiplexed click-based detectors for the first time. Applying the framework to real calibration data from three architecturally distinct detectors, we demonstrate direct, single-metric comparison across the photon-number regimes relevant to both DV and CV computing. Presupposing neither a detector principle nor an application, the framework generalizes prior single-detector analyses into one common, universally comparable metric.
\end{sloppypar}
\end{abstract}

\section*{Introduction}

Photon-number-resolving (PNR) detectors are a component common to essentially all approaches to photonic quantum information processing. Regardless of architecture, photonic quantum computing universally requires non-Gaussianity (nonlinearity) that Gaussian operations alone cannot supply~\cite{Mari2012,Bartlett2002}, and counting photons is the principal means by which a purely optical system can provide it. In continuous-variable (CV) architectures, PNR detectors generate the non-Gaussian resource states needed to synthesize Gottesman-Kitaev-Preskill (GKP) qubits~\cite{GKP2001}, which are then processed on large-scale, time-domain-multiplexed cluster states~\cite{Asavanant2019,Larsen2019}. In discrete-variable (DV) architectures, PNR detectors herald single photons and confirm photon number in fusion-based computation. In every case, a PNR detector acts as a single-shot discriminator: it conditions a downstream quantum state or operation on the outcome of just one measurement, and this determination must be made in real time~\cite{Morais2024}. How to evaluate PNR-detector performance is therefore a question that cuts across every architecture.

PNR-detector performance has conventionally been assessed by fitting a Gaussian to each photon-number peak in a pulse-height histogram accumulated over many shots, estimating a detection probability from the fit, and reporting a ``fidelity'' as the overlap between this estimate and a theoretical model that accounts only for detection-efficiency loss. However, separating the areas of overlapping peaks by fitting is a population-estimation problem: however much the peaks overlap, accumulating enough shots recovers the true detection probabilities to arbitrary accuracy. Single-shot photon-number discrimination is a fundamentally different problem -- an individual-event classification task that asks, for one already-obtained event, which true photon number it came from. The error probability of this single-event judgment is bounded below by the overlap of the peaks themselves and is not reduced by collecting more events. Conventional fidelity, which is built from a fit requiring many shots, therefore cannot in principle capture this single-shot misidentification.

More fundamentally, conventional fidelity is defined as the overlap between measured data and a theoretical model that accounts for detection-efficiency loss alone; finite energy resolution -- single-shot misidentification caused by the overlap of adjacent peaks -- enters neither the reference model nor the evaluation procedure. This paper first fills that gap. We quantify single-shot misidentification due to finite resolution as an explicit confusion matrix, kept structurally separate from the conventional (efficiency-only) confusion matrix, and compose the two to construct a single confusion matrix describing the single-shot response after \emph{both} sources of error. Keeping the two origins separate at the point of construction lets one diagnose how much of the infidelity is due to hardware efficiency and how much to resolution; recombining them then yields, as the diagonal element of the composed matrix, the probability that a single-shot measurement, given that truly $k$ photons were incident, reports $k$ -- which we adopt directly as the fidelity, and which applies unchanged whether a detector encodes photon number in a continuous energy response or in a discrete click pattern.

Plotting this diagonal element $F_k$ as a function of the heralded photon number $k$ directly visualizes how much fidelity is lost to detection-efficiency loss versus resolution-driven misidentification, and in which photon-number regime each dominates, yielding a concrete design target for the detection efficiency and resolution required at a given photon number. Detection-efficiency loss and resolution-driven misidentification are, moreover, two qualitatively different kinds of error. The former is a physical process in which a photon is genuinely lost; the latter is an informational error in which a photon is captured but its reported value is mistaken. Loss due to detection efficiency is fixed by the device and cannot be improved after the fact, whereas resolution-driven misidentification can be improved -- at the cost of measurement rate -- by narrowing the range of photon numbers a single shot is allowed to report.

Important prior work exists on evaluating PNR-detector performance itself. Morais \emph{et al.} demonstrated single-shot evaluation of a transition-edge sensor's (TES) photon-number-discrimination performance by actually performing single-shot measurements~\cite{Morais2024}. The advantage of the present framework is that it enables single-shot evaluation even without access to a single-shot measurement setup, or from data already on hand -- it is a generalization of that approach. Separately, Provazn\'{i}k \emph{et al.} showed how to translate the closeness of a multiplexed on/off-detector array (an MSPD, built from $M$ multiplexed click detectors) to an ideal PNR detector, in the practical setting of conditional photon-number-state generation, into an equivalent number $M$ of detectors that would need to be multiplexed to match that performance~\cite{Provaznik2020}. That metric fits naturally with the design of click-based multiplexed detectors. The single-shot fidelity $F_k$ defined here is numerically identical to the fidelity of the same conditional photon-number-state generation process, so the two rest on the same quantitative foundation. Building on that foundation, this paper addresses questions specific to detectors with a continuous output -- pulse height or energy -- namely how detection efficiency and resolution each degrade fidelity, and how far the effect of finite resolution can be recovered by adjusting the accept region.

The remainder of the paper proceeds as follows. We first formalize single-shot photon-number discrimination. We then construct a single-shot POVM that incorporates finite resolution and define fidelity as its diagonal element. We next show how detection efficiency and resolution each degrade fidelity as a function of heralded photon number, and how resolution-driven degradation can be recovered, at the cost of generation rate, by adjusting the accept region used for the discrimination. Finally, we apply this framework to real data from three architecturally distinct detectors -- a multiplexed superconducting nanowire single-photon detector (SNSPD), a rising-edge-timing SNSPD, and a TES -- and quantitatively compare, on a common metric, which detector is preferable in the photon-number regimes required by DV and CV architectures respectively.

\section*{Single-Shot Photon-Number Discrimination}

A quantum measurement is generally described by a positive operator-valued measure (POVM), a set of positive operators $\{\hat{\Pi}_m\}$. Each element satisfies $\hat{\Pi}_m \ge 0$ and completeness,
\begin{equation}
\sum_m \hat{\Pi}_m = \hat{I}.
\end{equation}
For an input state $\hat{\rho}$, the probability of obtaining outcome $m$ is $p(m) = \mathrm{Tr}[\hat{\rho}\,\hat{\Pi}_m]$. For a photon-number-resolving detector, outcome $m$ is interpreted as the reported photon number. An ideal PNR detector realizes the projective measurement $\hat{\Pi}_m^{\mathrm{ideal}} = |m\rangle\langle m|$, returning $m=n$ with unit probability given a Fock state $|n\rangle$. Because real detectors respond independently of phase, their POVM elements are diagonal in the Fock basis,
\begin{equation}
\hat{\Pi}_m = \sum_{n=0}^{\infty} P(m|n)\,|n\rangle\langle n|.
\end{equation}
Here $P(m|n)=\langle n|\hat{\Pi}_m|n\rangle$ is the conditional probability of reporting $m$ given that truly $n$ photons were incident; the matrix formed from these elements is the confusion matrix. Because every $n$ is necessarily assigned to some $m$, $\sum_m P(m|n) = 1$ for all $n$. An ideal detector corresponds to $P(m|n)=\delta_{mn}$, i.e., the identity matrix.

PNR-detector performance has conventionally been evaluated by fitting a Gaussian to each photon-number peak in a pulse-height histogram to obtain detection probabilities; the present work constructs its own conventional POVM the same way. We write the resulting measured confusion matrix as $P_{\mathrm{exp}}(m|k)$. This is a conditional detection probability averaged over an ensemble of many shots and, for the reasons given next, does not by itself adequately capture the single-shot misidentification structure.

Conventional ensemble-derived detection probabilities do not directly quantify single-shot misidentification. Estimating the weight of neighboring photon-number peaks from a set of many events $x_1,\dots,x_N$, via least squares or maximum likelihood, is a population-estimation problem. As long as the two Gaussian components $g_a(x)$ and $g_{a+1}(x)$ are linearly independent (i.e., as long as their centers satisfy $\mu_a\neq\mu_{a+1}$, however much the peaks overlap), this estimator is asymptotically unbiased and its standard error converges to zero as $N\to\infty$. In other words, however much the peaks overlap, accumulating more statistics lets one recover the true detection probabilities to arbitrary accuracy. Single-shot photon-number discrimination, by contrast, is an individual-event classification problem: given one already-obtained event $x$, decide which true photon number it came from. The error probability of this judgment is not improved at all by collecting more events. Population-estimation accuracy improves with the sample size $N$, while the difficulty of classifying an individual event does not depend on $N$ at all -- this contrast is precisely why single-shot misidentification must be evaluated directly.

One might wonder whether, instead of a Gaussian fit, a simple threshold rule -- placing a decision boundary between adjacent peaks and counting events on either side -- would suffice. This, too, fails to expose single-shot misidentification. Events leaking from the $n$-photon distribution into the $(n-1)$-photon bin can be numerically offset by events leaking from the $(n-1)$-photon distribution into the $n$-photon bin. In particular, if performance is evaluated using a light source whose detection probabilities for neighboring photon numbers do not differ greatly, and if each peak is close to symmetric, this cancellation can make the threshold-derived detection probabilities appear correct even though single-shot misassignment is substantial. Single-shot misidentification is therefore hidden from an ensemble-averaged detection probability in principle, regardless of whether it is obtained by fitting or by any other route.

\section*{Constructing the Single-Shot POVM}

PNR-detector fidelity has conventionally been defined by comparing the measured confusion matrix $P_{\mathrm{exp}}(m|k)$ against a theoretical confusion matrix that accounts only for detection efficiency $\eta$,
\begin{equation}
P_{\eta}(m|k) = \binom{k}{m}\eta^{m}(1-\eta)^{k-m}\qquad(m\le k),
\label{eq:binom-loss}
\end{equation}
and computing the overlap of the two distributions (e.g., the Bhattacharyya coefficient). This definition incorporates finite energy resolution -- single-shot misidentification caused by the overlap of adjacent peaks -- into neither the reference model nor the treatment of the measured data. $P_\eta$ is merely a theoretical model that assumes ideal resolution, and $P_{\mathrm{exp}}$ itself is typically already an area separated out by the fitting procedure described above, so this comparison never brings the effect of resolution into the evaluation in the first place.

We therefore begin by explicitly constructing single-shot misidentification due to finite resolution as its own, independent confusion matrix. We approximate the distribution of the output value (pulse height or energy) corresponding to $n$ photons by a Gaussian with mean $\mu_n$ and standard deviation $\sigma_n$,
\begin{equation}
g_n(x) = \frac{1}{\sqrt{2\pi}\,\sigma_n}\exp\!\left[-\frac{(x-\mu_n)^2}{2\sigma_n^2}\right].
\end{equation}
In many energy-resolving detectors, $\sigma_n$ grows with $n$ (for a TES, for example, the sensitivity of the superconducting transition curve falls off on the high-energy side), so the overlap between neighboring peaks grows with photon number.

We place a decision boundary $x_n$ between the peaks for adjacent photon numbers $n$ and $n+1$. In the simplest case this is the point of equal likelihood, $g_n(x_n)=g_{n+1}(x_n)$ (as discussed below, this boundary can also be adjusted), giving the $n$-th bin as $(x_{n-1},x_n)$. The probability that an event truly corresponding to photon number $a$ falls into the bin classified as $b$ is
\begin{equation}
R_{ab} = \int_{x_{b-1}}^{x_b} g_a(x)\,dx,
\label{eq:Rab}
\end{equation}
satisfying $\sum_b R_{ab}=1$ for each $a$. The diagonal element $R_{aa}$ is the probability that peak $a$ remains in its own bin, while the off-diagonal elements $R_{a,a\pm1},R_{a,a\pm2},\dots$ give the probability of leaking into neighboring bins, decreasing with distance. Because the overlap between adjacent response peaks typically decays rapidly in the tails, components with large $|a-b|$ are negligible in practice.

The Gaussian form of $g_n(x)$ above is a convenient choice that matches our calibration data well, not a requirement of the construction: Eq.~(\ref{eq:Rab}) needs only some response distribution whose tail areas can be evaluated over each bin, whatever its shape -- the construction and the resulting confusion matrix $R_{ab}$ carry over unchanged.

\begin{figure}[htbp]
  \centering
  \includegraphics[width=1\linewidth]{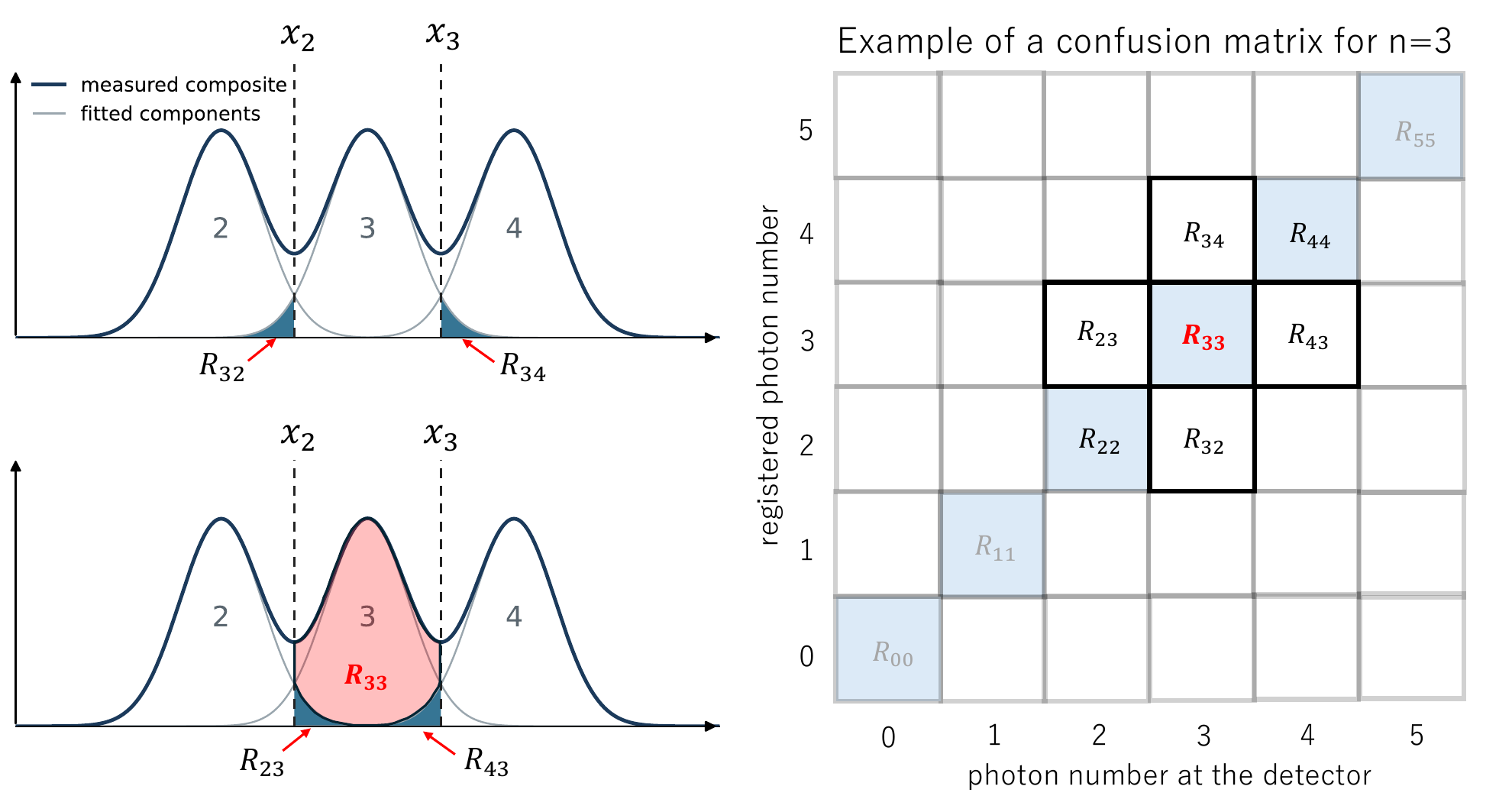}
  \caption{A concrete example of the resolution confusion matrix $R_{ab}$, centered on $a=3$. The horizontal axis is the true photon number and the vertical axis is the reported photon number. The diagonal element $R_{33}$ carries the dominant weight, with leakage into neighboring photon numbers, $R_{32}$ and $R_{34}$, appearing as off-diagonal elements. This matrix represents single-shot misidentification arising from finite energy resolution alone.}
  \label{fig:confusion_def}
\end{figure}

Figure~\ref{fig:confusion_def} shows a concrete example of the confusion matrix $R_{ab}$ centered on $a=3$. The diagonal element $R_{33}$ dominates, but overlap with neighboring bins produces off-diagonal elements such as $R_{32}$ and $R_{34}$, with the more distant $R_{31}$ and $R_{35}$ appearing at even smaller values. This matrix $R_{ab}$ is precisely the quantity representing single-shot misidentification arising from finite energy resolution alone.

\begin{figure}[htbp]
  \centering
  \includegraphics[width=1\linewidth]{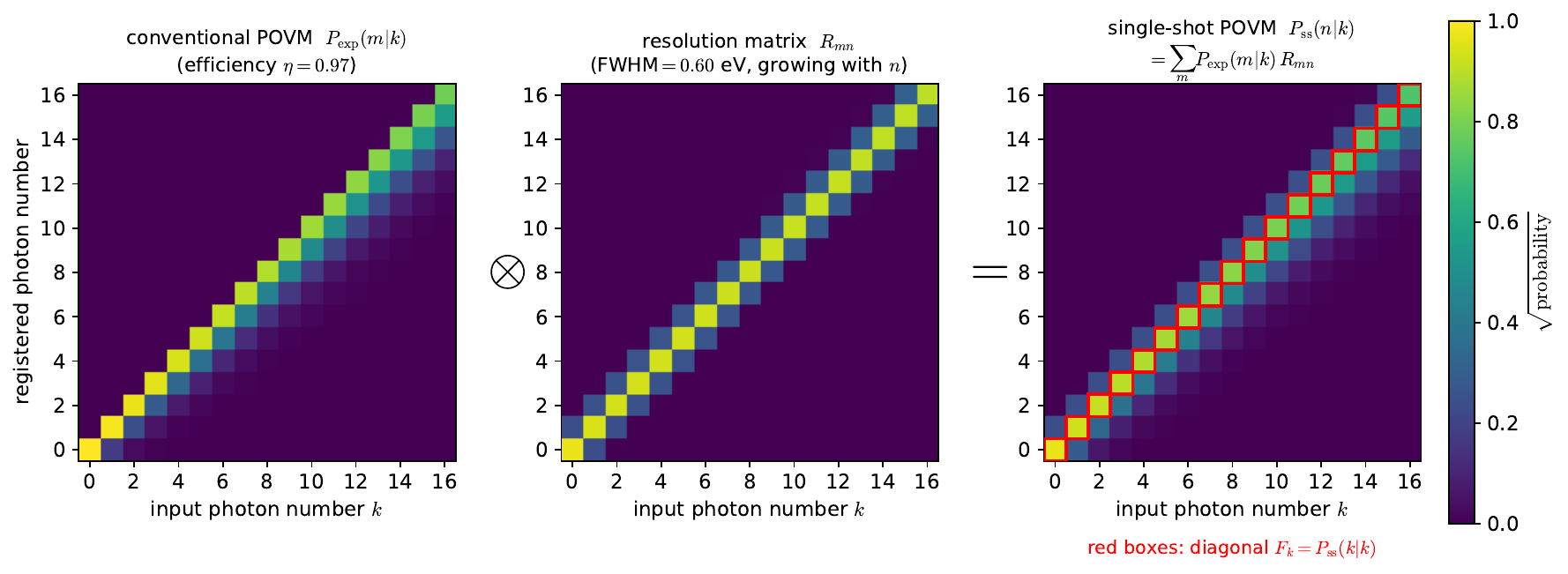}
  \caption{Construction of the single-shot POVM. Shown here for illustration is a $1550\,\mathrm{nm}$ photon measurement with detection efficiency $97\,\%$ and resolution $0.6\,\mathrm{eV}$. Composing the conventional POVM $P_{\mathrm{exp}}(m|k)$, which includes only the detection efficiency ($\eta$) and small residual label errors (left), with the resolution confusion matrix $R_{mn}$ defined in Fig.~\ref{fig:confusion_def} (center), yields the single-shot POVM $P_{\mathrm{ss}}(n|k)=\sum_m P_{\mathrm{exp}}(m|k)R_{mn}$ (right), which carries single-shot misidentification due to the overlap of neighboring peaks as off-diagonal elements.}
  \label{fig:povm_convolution}
\end{figure}

Composing $R_{ab}$ with the measured confusion matrix $P_{\mathrm{exp}}(m|k)$ -- which includes detection efficiency together with the small residual label errors that an ordinary fit can capture -- yields the confusion matrix for single-shot evaluation,
\begin{equation}
P_{\mathrm{ss}}(n|k) = \sum_m P_{\mathrm{exp}}(m|k)\,R_{mn}
\label{eq:ss}
\end{equation}
(Fig.~\ref{fig:povm_convolution}). Both $P_{\mathrm{exp}}(m|k)$ and $R_{mn}$ are row-stochastic matrices, and Eq.~(\ref{eq:ss}) is their matrix product. This composition links two sequential transitions -- the true photon number $k$ becomes an intermediate reported value $m$ through efficiency loss, and $m$ then becomes the final reported value $n$ through resolution-driven misidentification -- and is nothing other than the Chapman-Kolmogorov equation for a two-step Markov chain. The Markov property at work here, in which the misidentification probability $R_{mn}$ from $m$ to $n$ depends only on the intermediate value $m$ and not on the earlier value $k$, is not a mere mathematical convenience: it is physically required by the causal structure of the measurement -- efficiency loss occurs upstream of the detector, and the energy-resolved response is determined solely by the photon number $m$ that actually reaches the detector. The factorization assumes that the detector response depends only on the photon number that actually reaches the detector, and not on the history by which that number was obtained. $P_{\mathrm{ss}}(n|k)$ is thus the probability that a single shot reports $n$, given that truly $k$ photons were incident, after passing through both detection-efficiency loss and finite-resolution single-shot misidentification.

The question that remains is what $P_{\mathrm{ss}}(n|k)$, once constructed this way, should be compared against to obtain a fidelity. Following convention, one would compare $P_{\mathrm{ss}}$ against some theoretical model such as Eq.~(\ref{eq:binom-loss}) and compute their overlap. But that theoretical model $P_\eta$ is merely a convenient reference point -- ``ideal except for detection-efficiency loss.'' What this paper actually wants to ask is a different and more direct question. When a PNR detector is used in photonic quantum computing, what is expected of it is that a single shot deliver the correct Fock state $|k\rangle$, not that it match some particular loss model. There is therefore no need to invoke a theoretical model $P_\eta$ as a point of comparison at all: one need only read off, directly, the $n=k$ component of $P_{\mathrm{ss}}(n|k)$ -- the probability that a single-shot measurement, given that truly $k$ photons were incident, correctly reports $k$. We accordingly define the single-shot fidelity as
\begin{equation}
F_k \;:=\; P_{\mathrm{ss}}(k|k),
\label{eq:fid}
\end{equation}
the diagonal element of the confusion matrix constructed in Eq.~(\ref{eq:ss}), which requires no separate reference theoretical model. Once the confusion matrix has been constructed, reading off its diagonal yields the fidelity $F_k$ arising from both detection efficiency and resolution together. In what follows, we examine $F_k$ as a function of the heralded photon number $k$ to reveal how detection efficiency and resolution each affect fidelity.

\section*{How Detection Efficiency and Resolution Degrade Fidelity}

We proceed below using a TES as an example energy-resolving PNR detector. The quantity $F_k$ defined in Eq.~(\ref{eq:fid}) represents how reliable a single-shot measurement heralding photon number $k$ actually is. Plotting it as a function of the heralded photon number $k$ directly visualizes which photon-number regime loses how much fidelity to detection-efficiency loss versus resolution-driven misidentification, revealing which of the two origins needs to be addressed and how. The effect of detection efficiency $\eta$ follows directly from Eq.~(\ref{eq:binom-loss}): under ideal resolution, $F_k=\eta^k$, and however close $\eta$ is to unity, $F_k$ falls off exponentially as $k$ grows. Even at $\eta=0.99$, $F_k$ drops to about $0.85$ by $k=16$. Because of this exponential sensitivity, a small drop in detection efficiency costs a great deal of fidelity, making high detection efficiency a basic prerequisite for high fidelity. Detection efficiency is intrinsic to the device, and because it reflects genuine photon loss, there is no way to recover it after the fact; keeping detection efficiency sufficiently high is therefore a non-negotiable requirement. The effect of resolution, captured by the diagonal element $R_{kk}$ of Eq.~(\ref{eq:Rab}), is set by the degree of overlap with neighboring peaks. Examining real TES calibration data, we find that $\sigma_n$ grows gently (nearly linearly) with $n$, while the peak spacing $\mu_{n+1}-\mu_n$ itself also shrinks slightly (plausibly a consequence of response compression near the top of the superconducting transition curve); together these cause $R_{kk}$ to decrease with $k$.

\begin{figure}[htbp]
  \centering
  \includegraphics[width=1\linewidth]{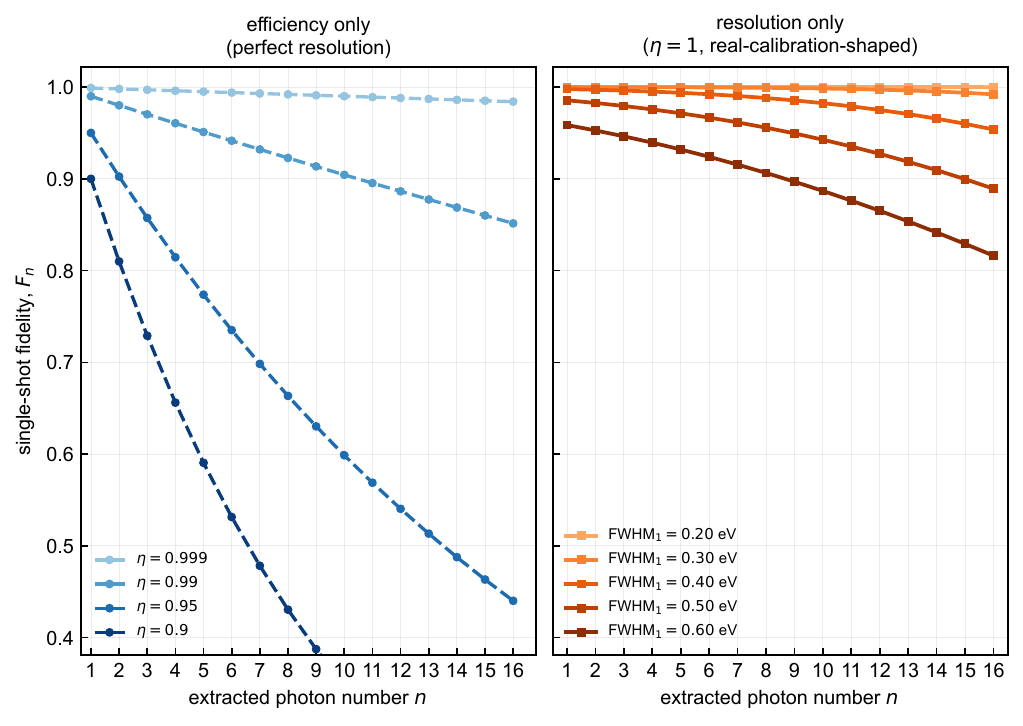}
  \caption{Dependence of single-shot fidelity $F_k$ on the heralded photon number $k$, for a $1550\,\mathrm{nm}$ photon measurement. Horizontal axis: heralded photon number $k$; vertical axis: $F_k$ (linear scale). (a) Energy resolution fixed, detection efficiency $\eta$ varied (e.g., $\eta=1.0,0.99,0.95,0.90$). (b) Detection efficiency fixed at $\eta=1$, energy resolution FWHM varied over $0.20,0.30,0.40,0.50,0.60\,\mathrm{eV}$. Overlaying the two series shows that detection efficiency and resolution each independently degrade $F_k$.}
  \label{fig:fidelity_vs_n}
\end{figure}

Figure~\ref{fig:fidelity_vs_n} plots $F_k$ as a function of the heralded photon number $k$ for several values of detection efficiency and several values of energy resolution. The resolution series uses a calibrated model that preserves the growth in $\sigma_n$ and the compression in peak spacing obtained from real device calibration data, varying only the FWHM. The most important point this figure makes is that degradation from resolution alone is of the same order of magnitude as degradation from detection efficiency alone. Resolution-driven misidentification is therefore never a secondary effect that can be neglected once detection efficiency is secured. The right way to think about these two origins can accordingly be summarized as follows. Detection efficiency $\eta$ is a constraint intrinsic to the detector hardware, with no room for adjustment beyond the form $\eta^k$; it is a prerequisite. Pursuing high fidelity must begin by securing sufficient detection efficiency. But even with sufficient detection efficiency secured, Fig.~\ref{fig:fidelity_vs_n} shows that resolution-driven degradation is never negligible in size, and leaving it unaddressed prevents reaching fidelities above roughly $90\,\%$. Unlike detection efficiency, however, the resolution-driven term $R_{kk}$ \emph{can} be improved -- at the cost of generation rate -- by adjusting the accept region used for the discrimination. The prerequisite of detection efficiency once cleared, resolution is thus the remaining, genuinely addressable challenge, and the next section discusses how far, and at what cost, fidelity can be improved.

\section*{Trade-off between Fidelity and Generation Rate via the Accept Region}
\label{sec:tradeoff}

The fidelity loss from resolution seen in the previous section can be adjusted by changing the strictness of the discrimination. This corresponds, in signal detection theory, to the separation between sensitivity (a property intrinsic to the detector) and criterion (a threshold that can be chosen at run time)~\cite{GreenSwets1966}. Up to this point we have implicitly placed the decision boundaries $\{x_m\}$ of Eq.~(\ref{eq:Rab}) at the ordinary equal-likelihood points, which allow no rejection; moving these boundaries changes the effect of resolution itself.

\begin{figure}[htbp]
  \centering
  \includegraphics[width=1\linewidth]{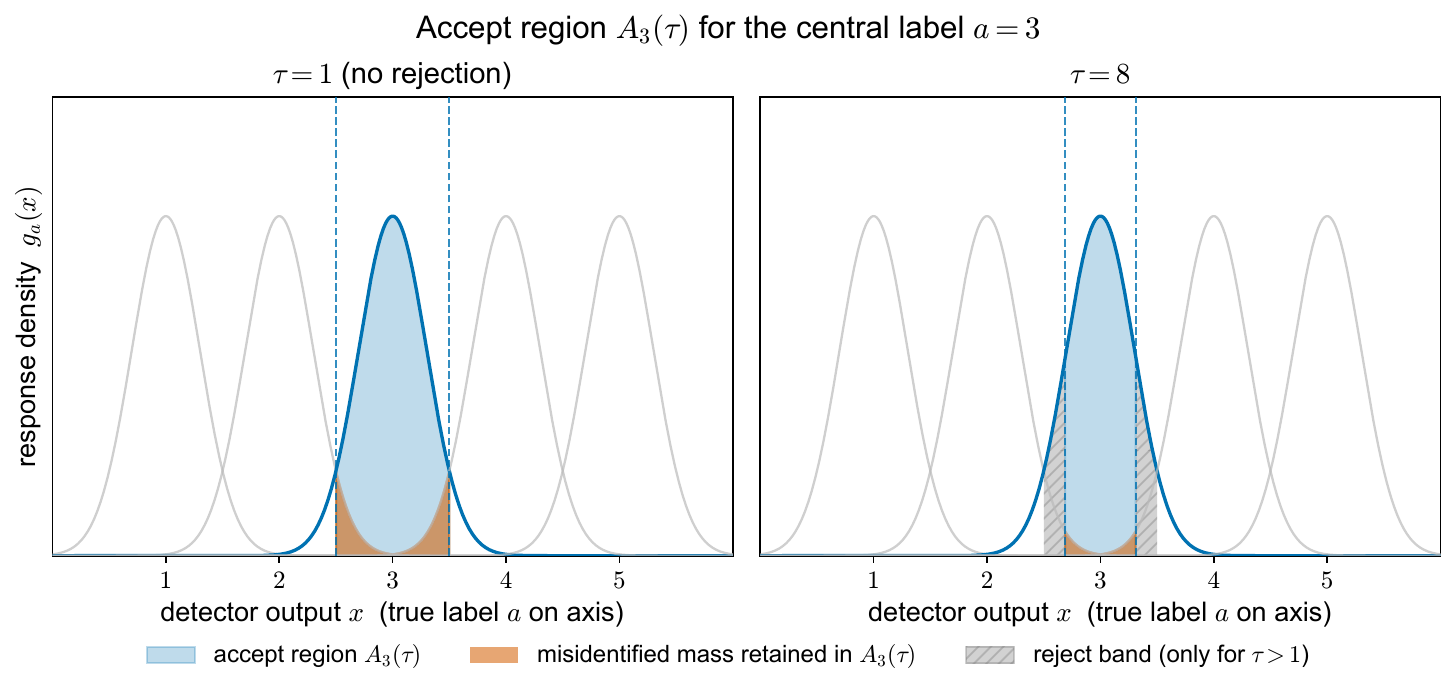}
  \caption{Schematic of the accept region. For the confusion matrix centered on $a=3$, the accept regions for $\tau=1$ (no rejection, the ordinary equal-likelihood boundary) and for $\tau>1$ are shown side by side. For $\tau>1$, reject bands appear near the boundaries with neighboring peaks; the figure shows both the contraction of the accept region toward the center and the fraction of events falling into the reject bands.}
  \label{fig:reject}
\end{figure}

Concretely, we introduce the option of rejecting, as ``undetermined,'' any event about which the discrimination is not confident. This generalizes unambiguous discrimination in quantum measurement theory -- a measurement strategy that permits no erroneous determination in exchange for allowing inconclusive outcomes~\cite{Ivanovic1987,Dieks1988,Peres1988,Chefles2000} -- into a form that can be tuned continuously via a threshold. We introduce a threshold $\tau\,(\ge1)$ such that the true photon number $a$ is reported as ``$a$'' only if the likelihood ratio against a neighboring photon number $b$ satisfies
\begin{equation}
\frac{g_a(x)}{g_b(x)} \ge \tau,
\end{equation}
and any event that fails this condition against every neighboring candidate is rejected. Equation~(\ref{eq:Rab}) then generalizes to
\begin{equation}
R_{a,n}(\tau) = \int_{A_n(\tau)} g_a(x)\,dx,
\label{eq:Ran-tau}
\end{equation}
where $A_n(\tau)$ is the accept region determined by the likelihood-ratio condition above, and $\tau=1$ recovers the ordinary equal-likelihood boundary of Eq.~(\ref{eq:Rab}) with no rejection. As $\tau$ increases, the accept region contracts toward the center $\mu_a$; the fraction of events rejected,
\begin{equation}
P_{\mathrm{reject}}(a;\tau) = 1-\sum_n R_{a,n}(\tau),
\end{equation}
grows, while the probability that an accepted event is misjudged shrinks (Fig.~\ref{fig:reject}).

This operation is meaningful for the following reason. Resolution-driven misidentification leaves, in the output value $x$ itself used for the discrimination, information about how confident that discrimination actually is. Rejecting events about which one is not confident can therefore, in principle, raise accuracy without limit. In other words, resolution-driven misidentification can be converted, by tightening the discrimination, into an erasure -- an error for which one knows at the time of judgment which trials are untrustworthy. In the context of quantum error correction, converting errors into erasures is known to substantially raise the fault-tolerance threshold~\cite{Wu2022,Kubica2023}, and detectable loss can be tolerated at a far higher rate than undetectable error~\cite{Barrett2010}. The present discussion applies this general principle to the specific context of single-shot photon-number discrimination.

Detection-efficiency loss, by contrast, offers no such recourse. Consider an event in which truly $k$ photons were incident but one was lost before detection. The output value $x$ observed for this event is statistically identical to the output-value distribution of an event in which truly $k-1$ photons were incident and detected without loss; nothing in $x$ retains any trace that this event was originally $k$ photons. Loss is a process that rewrites the photon number itself prior to the measurement, and no amount of tightening the discrimination after the fact can detect or reject it. This is why we stated in the previous section that there is no room for adjustment beyond the form $\eta^k$.

Because of this asymmetry, increasing $\tau$ is a pure trade-off: one sacrifices the fraction of events accepted in exchange for accepting only events one can be confident about. Let the probability that a truly-$n$-photon event is accepted be
\begin{equation}
Y_n(\tau) = \sum_b R_{n,b}(\tau).
\end{equation}
$Y_n(\tau)$ decreases monotonically with $\tau$. Restricting to accepted events, define the conditional fidelity
\begin{equation}
C_n(\tau) = \frac{R_{n,n}(\tau)}{Y_n(\tau)},
\end{equation}
which represents the quality of output an application actually receives, excluding rejected events from the population from the outset. At $\tau=1$ no rejection occurs, so $Y_n(1)=1$ and $C_n(1)=R_{n,n}(1)=F_n$, recovering exactly the definition of the previous section. As $\tau$ increases, $C_n(\tau)$ improves -- but not monotonically all the way to $1$. In the calibrated model where $\sigma_n$ grows with $n$, the next-lower neighboring peak ($n-1$) is always narrower than peak $n$ itself, so raising $\tau$ too far causes the accept region on that side to vanish entirely at a finite $\tau^\ast$, beyond which $C_n(\tau)$ actually worsens. There is therefore an optimum $\tau^\ast=\arg\max_\tau C_n(\tau)$, set by the resolution, beyond which further tightening serves no purpose.

\begin{figure}[htbp]
  \centering
  \includegraphics[width=0.85\linewidth]{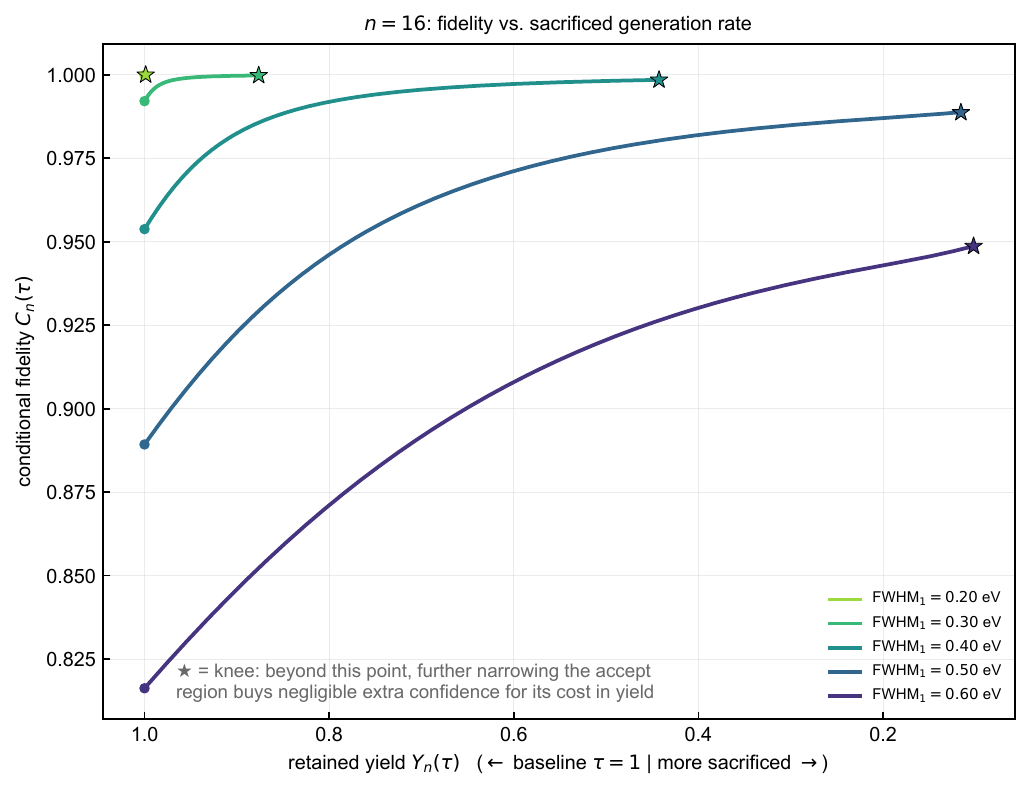}
  \caption{Trade-off between fidelity and generation rate via the accept region. For a fixed target photon number $n$, sweeping the threshold $\tau$ from $1$ to $\tau^\ast$ traces out the yield $Y_n(\tau)$ (horizontal axis, corresponding to the generation rate sacrificed) against the conditional fidelity $C_n(\tau)$ (vertical axis). Curves are shown for several values of energy resolution. Each curve terminates at the optimum point $\tau^\ast$ beyond which raising $\tau$ no longer helps (star). Detectors with better resolution reach higher $C_n$, and sacrifice less generation rate to reach a given $C_n$.}
  \label{fig:rate-tradeoff}
\end{figure}

Multiplying this yield by the repetition rate $R_{\mathrm{trial}}$ at which the detector-and-source system can generate trials per unit time gives the number of confident reports of photon number $n$ per unit time, the confident generation rate,
\begin{equation}
\Gamma_n(\tau) = R_{\mathrm{trial}}\times Y_n(\tau)\quad[\mathrm{Hz}].
\end{equation}
Figure~\ref{fig:rate-tradeoff} shows, for a fixed target photon number $n$, how much the conditional fidelity $C_n(\tau)$ improves in exchange for the generation rate sacrificed (the decrease in yield $Y_n(\tau)$) as the threshold $\tau$ is swept from $1$ to $\tau^\ast$. Even a detector with poor resolution can, by adopting a narrow accept region while operating fast enough (i.e., with sufficiently large $R_{\mathrm{trial}}$), sometimes compensate for the fidelity loss due to insufficient resolution shown in Fig.~\ref{fig:fidelity_vs_n} while maintaining a practical generation rate. Note, however, that this improvement is bounded, with the bound set by $\tau^\ast$.

The degradation shown by the resolution series in Fig.~\ref{fig:fidelity_vs_n} is thus not fixed: it can be improved, in exchange for generation rate, along the curves of Fig.~\ref{fig:rate-tradeoff} (bounded above by $\tau^\ast$) by adjusting the accept region. $\tau$ is an operational parameter to be chosen according to the confidence an application demands, not a fixed property of the detector itself.

\section*{Independence from Detector Operating Principle}

This framework requires only that a detector's output retain some gradient of confidence about the true photon number. A detector such as a TES, which outputs a continuous pulse height or energy, satisfies this condition and admits a reject option via adjustment of the accept region. A detector based on multiplexing many on/off elements (an MSPD), by contrast, has already compressed its determination into a single integer, and this room for rejection is absent in principle. In an MSPD, for instance, the simultaneous incidence of multiple photons on the same element (a collision) is reflected only in a binary click/no-click output, with no information surviving in the output about whether a given event involved a collision; there is consequently no way to recover accuracy after the fact by tightening the discrimination. Errors arising in an MSPD therefore appear, like detection-efficiency loss, as an unrecoverable performance limit. This distinction is not a weakness of any particular detector, but an essential feature of single-shot evaluation intrinsic to the architecture itself.

\section*{Interim Summary}

This paper incorporates single-shot misidentification due to finite resolution, expressed as its own confusion matrix, into the measured POVM, keeping it structurally separate from detection efficiency at the point of construction. Composing the two together and reading off the diagonal element $F_k=P_{\mathrm{ss}}(k|k)$ of the result yields a single-shot fidelity that isolates how much performance is lost to each origin, while remaining directly comparable across detectors regardless of whether photon number is encoded in a continuous energy response or a discrete click pattern. This quantity coincides with the fidelity of the standard practical scenario of conditional photon-number-state generation.

Examining $F_k$ as a function of the heralded photon number $k$ directly visualizes how much fidelity is lost to detection efficiency versus resolution in each photon-number regime, and reveals the detection efficiency and resolution required for a given target photon number. We further showed that these two origins are asymmetric. Detection-efficiency loss offers no means of converting a judgment into an erasure, and stands as an unrecoverable performance limit, whereas resolution-driven misidentification can be improved -- at the cost of generation rate -- by adjusting the accept region used for discrimination. Building on this asymmetry, we formalized an explicit trade-off, using the pair $Y_n(\tau)$ and $C_n(\tau)$, in which narrowing the accept region raises fidelity at the cost of generation rate. This improvement does not continue indefinitely, however; we also showed that it is bounded by an optimum $\tau^\ast$ specific to each resolution.

The following section applies this framework to real data, comparing it against detectors of other architectures and connecting it to the performance levels demanded by the leading approaches to photonic quantum computing.

\section*{Cross-Detector Evaluation with Real Data}

We apply the framework developed above to three PNR-detector architectures with published characterization data: multiplexed SNSPDs, rising-edge-timing SNSPDs, and TESs. Although DV and CV architectures demand qualitatively different photon-number-discrimination capabilities, we show concretely, using real data, that both can be evaluated on the same metric, the single-shot fidelity $F_k=P_{\mathrm{ss}}(k|k)$.

\subsection*{Photon-Number Discrimination Requirements of DV and CV Architectures}

\begin{sloppypar}
In DV architectures (such as fusion-based photonic computing), the resource states are small entangled states of only a few photons, so the required photon-number-discrimination capability is confined to the low-photon-number regime $n=0$--$4$~\cite{KLM2001,Bartolucci2023}, and real PNR detectors are correspondingly designed with this regime as their primary performance target~\cite{PsiQuantum2025}.
\end{sloppypar}

CV architectures, typified by GKP breeding~\cite{Takase2023}, demand discrimination at substantially higher photon numbers ($n\sim16$). It is worth first characterizing the distribution of the true photon number actually incident on the detector. The probability distribution of the detected photon number in Takase \emph{et al.}'s generalized photon-subtraction process is
\begin{equation}
P(n) \propto \binom{2n}{n}4^{-n}\left(\frac{t}{t+2}\right)^{n},
\end{equation}
where the parameters correspond to a target photon number of $n=16$~\cite{Takase2023}. This distribution is smoothly decreasing, close to that of a thermal state, with no parity gap. The prior probability of an event that was truly $15$ or $17$ photons is comparable to, or greater than, that of the target $16$ photons.

Under this distribution, photon loss due to detection efficiency (a one-directional, binomially distributed process that removes photons from a higher true photon number) and misidentification due to finite energy resolution (a Gaussian-like process that falls off as the square of the distance) reach very different ranges of error. What resolution must actually discriminate against is therefore \textbf{neighboring} photon numbers ($15$ and $17$, and, weakly, $14$ and $18$); misidentification against more distant photon numbers appears as a detection-efficiency problem rather than a resolution problem. With this in mind, we use $F_{16}$, which combines detection efficiency and resolution, as the representative evaluation metric for CV architectures below.

\begin{figure}[htbp]
  \centering
  \includegraphics[width=1\linewidth]{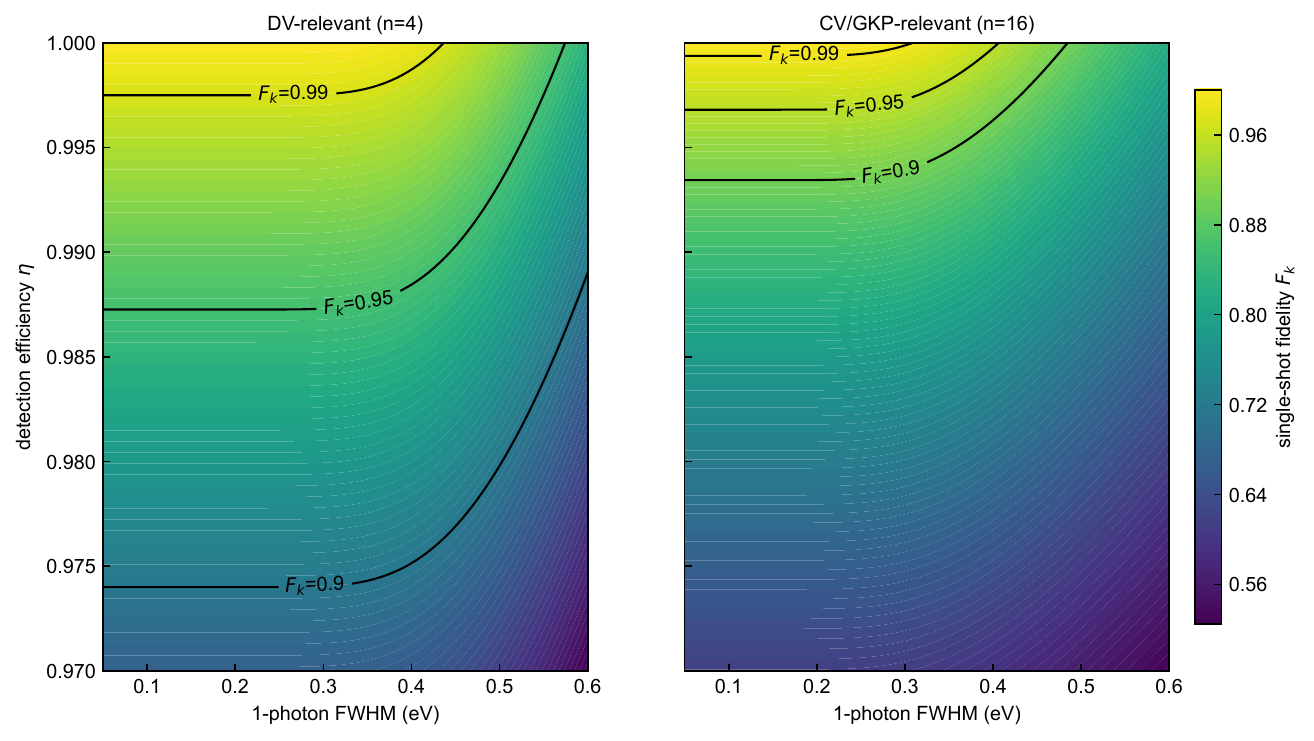}
  \caption{Effect of detection efficiency $\eta$ and resolution (FWHM of the one-photon peak) on single-shot fidelity $F_k$, shown as fidelity contours in the $(\mathrm{FWHM}_1,\eta)$ plane. (a) $n=4$ (the photon number relevant to DV architectures). (b) $n=16$ (the photon number relevant to CV architectures and GKP breeding). Color indicates the value of $F_k$ (see color bar); lines mark the principal contours ($F_n=0.9,0.95,0.99$). The shape of the resolution model (the growth of $\sigma_n$ and the compression of peak spacing obtained from real device calibration data) is held fixed while only the FWHM is varied. At $n=4$, resolution of $\mathrm{FWHM}_1\lesssim0.4\,\mathrm{eV}$ combined with $\eta\gtrsim99.8\,\%$ reaches $F_4=0.99$; at $n=16$, reaching the same $F_{16}=0.99$ instead requires $\mathrm{FWHM}_1\lesssim0.2\,\mathrm{eV}$ and $\eta\gtrsim99.9\,\%$ simultaneously, visually confirming that CV architectures demand far more stringent performance than DV architectures.}
  \label{fig:fig2_efficiency_resolution_map}
\end{figure}

Figure~\ref{fig:fig2_efficiency_resolution_map} quantifies how differently detection efficiency and resolution must perform for the DV-relevant $n=4$ and the CV-relevant $n=16$. It shows contours of single-shot fidelity $F_n$ in the $(\mathrm{FWHM}_1,\eta)$ plane for $n=4$ (panel a) and $n=16$ (panel b). The range of $(\mathrm{FWHM}_1,\eta)$ combinations achieving $F_n=0.99$ -- a ``passing region'' -- shrinks dramatically as $n$ grows. At $n=4$, a resolution of $\mathrm{FWHM}_1\lesssim0.4\,\mathrm{eV}$ suffices to reach $F_4=0.99$ as long as $\eta\gtrsim99.8\,\%$. At $n=16$, by contrast, reaching the same $F_{16}=0.99$ requires \emph{simultaneously} $\mathrm{FWHM}_1\lesssim0.2\,\mathrm{eV}$ and $\eta\gtrsim99.9\,\%$; once $\mathrm{FWHM}_1$ is even one step worse (around $0.4\,\mathrm{eV}$), $F_{16}=0.99$ becomes unreachable in principle no matter how far detection efficiency is raised within this range. In short, DV architectures relax the requirement on resolution once detection efficiency is secured, whereas CV architectures require both detection efficiency and resolution to be simultaneously excellent -- a far more demanding requirement. The cross-detector comparison below is made with this asymmetric requirement in mind.

It is worth noting that not all CV applications share the same requirement. Several routes to GKP synthesis besides breeding have been proposed: a backcasting-search-based approach~\cite{Fukui2022} in which each detector registers at most two photons, an approach using approximate GKP codes~\cite{Tzitrin2020}, and larger-scale fault-tolerant designs~\cite{Bourassa2021}, each demanding a different photon-number range. Even the reason a high photon number is required in the first place traces back to the fault-tolerance threshold of the GKP code~\cite{Fukui2018,Noh2022}, so the required level itself depends on the target error-correction performance. Applications such as verification via Gaussian boson sampling~\cite{Zhong2020,Madsen2022} also exist, which lack breeding's parity structure and instead require accurately verifying an arbitrary photon-number distribution. The level $n\sim16$ evaluated below is thus specific to the concrete application of breeding-type GKP synthesis and should not be taken as a uniform requirement across all of CV computing.

\subsection*{Evaluation of Multiplexed SNSPDs}

In a multiplexed SNSPD (MSPD), misidentification arising from the simultaneous incidence of multiple photons on the same element (a collision) can only ever cause photon number to be underestimated~\cite{Paul1996,Provaznik2020}. This multiplexed architecture has developed in various forms, including designs optimized for high detection efficiency~\cite{Reddy2020} and for multi-photon discrimination on a single waveguide~\cite{Cahall2017,Zhu2020}. The upper bound $F_k\le\eta^k$ (Eq.~(\ref{eq:binom-loss})) given by an ideal PNR detector with detection efficiency $\eta$ alone therefore also holds for an MSPD in the limit $M\to\infty$ of infinite multiplexing -- an absolute ceiling that cannot be exceeded no matter how large $M$ is made.

Ding \emph{et al.} report a multiplexed SNSPD with detection efficiency $98\,\%$ and $M=32$~\cite{Ding2025}. This work is oriented toward DV architectures and general quantum-metrology applications, with no mention of CV-architecture GKP synthesis. The single-shot-equivalent fidelities measured by that paper's detector tomography are $97.5\,\%$ at one photon, approximately $87\,\%$ at two photons, $73\,\%$ at three photons, and $40\,\%$ at four photons -- already substantially degraded by $n=4$. A larger, $M=100$ multiplexed SNSPD has also been reported~\cite{Cheng2023}, but it, too, is not oriented toward operation at the $n\sim16$ scale. There is accordingly no need to extrapolate this trend out to $n=16$ for the comparison; the theoretical upper bound alone suffices. At $\eta=98\,\%$, $F_{16}\le\eta^{16}\approx72.4\,\%$, and this bound cannot be exceeded no matter how far $M$ is increased (even reaching $90\,\%$ of the bound for $F_{16}$ would require $M\gtrsim1100$).

\subsection*{Evaluation of Rising-Edge-Timing SNSPDs}

For SNSPDs that discriminate photon number via rising-edge-timing analysis, Schapeler \emph{et al.} reconstruct a POVM $\Pi$ for an SNSPD with detection efficiency $\eta=91\,\%$ using quantum detector tomography across many input mean photon numbers, fitting rising-edge-time histograms to exponentially modified Gaussian (EMG) distributions~\cite{Schapeler2026,Sidorova2025,Nicolich2019,Lundeen2009}. However, this reconstruction uses only the count of events within each fixed threshold region, not the continuous shape of the underlying response distribution -- the same simplification that, in a separate part of their analysis, Schapeler \emph{et al.} find ``significantly underestimates'' misidentification when a Gaussian is used in place of a more accurate exponentially modified Gaussian model. Whether their POVM reconstruction is subject to an analogous underestimation is not established, and cannot be ruled out. Their measured values, in a configuration restricting resolution to $n\le3$, are
\begin{equation}
F_1 = 87.1\,\%,\qquad F_2 = 74.8\,\%,\qquad F_3 = 65.4\,\%,
\end{equation}
and this detector saturates around $n\approx6$--$7$, beyond which it has not been characterized~\cite{Schapeler2026}. Unlike a multiplexed SNSPD, this saturation is not a fundamental limit of the kind imposed by the detection-efficiency-only ceiling; as the next subsection shows, it can be partially improved by adjusting the accept region. That said, the currently reported measured data do not demonstrate reaching the high-photon-number regime required by CV architectures.

\subsection*{Evaluation of the TES}

We evaluated a TES with system detection efficiency $\eta=99\,\%$~\cite{Jodoi2026} using the method proposed in this paper: detection probabilities are obtained by Gaussian fitting, a conventional POVM $P_{\mathrm{exp}}$ is reconstructed from these by maximum likelihood across the real calibration data, and this is composed with the resolution confusion matrix $R$ built from the same fits, following exactly the two-step construction of Eq.~(\ref{eq:ss}). Figure~\ref{fig:detector_comparison} overlays, with the other detectors, $F_k$ evaluated via Eq.~(\ref{eq:fid}) across the full range $k=1$ to $16$. $F_k$ falls from $F_1=94.2\,\%$ to $F_{16}=76.1\,\%$, not monotonically -- it dips to $74.5\,\%$ at $k=15$ before recovering, a fluctuation we attribute to statistical scatter in the real calibration data rather than to any physical mechanism. Because our construction keeps the efficiency and resolution contributions separate, we can directly attribute this shortfall: the efficiency-only ceiling at the same $\eta=99\,\%$ is $\eta^{16}\approx85.1\,\%$, so resolution alone accounts for roughly $9$ percentage points of additional fidelity loss at $k=16$ -- a gap that widens with photon number, consistent with the resolution-driven degradation identified in Fig.~\ref{fig:fidelity_vs_n}.

\subsection*{A Worked Example of the Accept-Region Trade-off}

All of the results above are for the unrestricted accept region ($\tau=1$); we now examine how far the trade-off between yield $Y_n(\tau)$ and conditional fidelity $C_n(\tau)$ from adjusting the accept region can raise these baselines. This trade-off is independently corroborated by two real datasets.

Schapeler \emph{et al.} show directly, from measured data, that narrowing the decision region for one-photon events -- at the cost of $6\,\%$ of the yield -- improves the misidentification probability roughly $14$-fold, from $0.14\,\%$ to $0.01\,\%$~\cite{Schapeler2026}. Although this is based on a definition strictly different from the $Y_n(\tau)$ and $C_n(\tau)$ used here, it is an empirical demonstration, qualitatively consistent with the theoretical claim of this paper, that fidelity can indeed be improved in exchange for yield by narrowing the accept region.

Performing the same calculation on our TES data, at $n=1$ the misidentification probability of $1.61\,\%$ at $\tau=1$ (no rejection) -- i.e., the resolution-driven component of $F_1=94.2\,\%$ above -- improves roughly $16$-fold, to $0.10\,\%$, at the cost of $14.9\,\%$ of the yield. This effect is even more pronounced at $n=16$: the misidentification probability of $10.36\,\%$ at $\tau=1$ improves roughly $8$-fold, to $1.27\,\%$, at the cost of $68.3\,\%$ of the yield. Here, the ``reachable level'' is a calculated value based on real calibration data at the point where further tightening $\tau$ no longer improves fidelity and only continues to sacrifice yield (the optimum $\tau^\ast$ introduced in the section on the accept-region trade-off) -- it is not a value obtained by actually re-measuring with the accept region narrowed. That the improvement from adjusting the accept region grows with photon number is consistent with the discussion above: unlike detection-efficiency loss (unrecoverable), resolution-driven misidentification (adjustable via the accept region) becomes more dominant at high photon number. This improvement, from baseline to the level reached after sacrificing yield, independently confirmed for two architecturally distinct detectors -- the TES and the rising-edge-timing SNSPD -- is also summarized in Fig.~\ref{fig:detector_comparison}.

\subsection*{Summary}

\begin{figure}[htbp]
  \centering
  \includegraphics[width=1\linewidth]{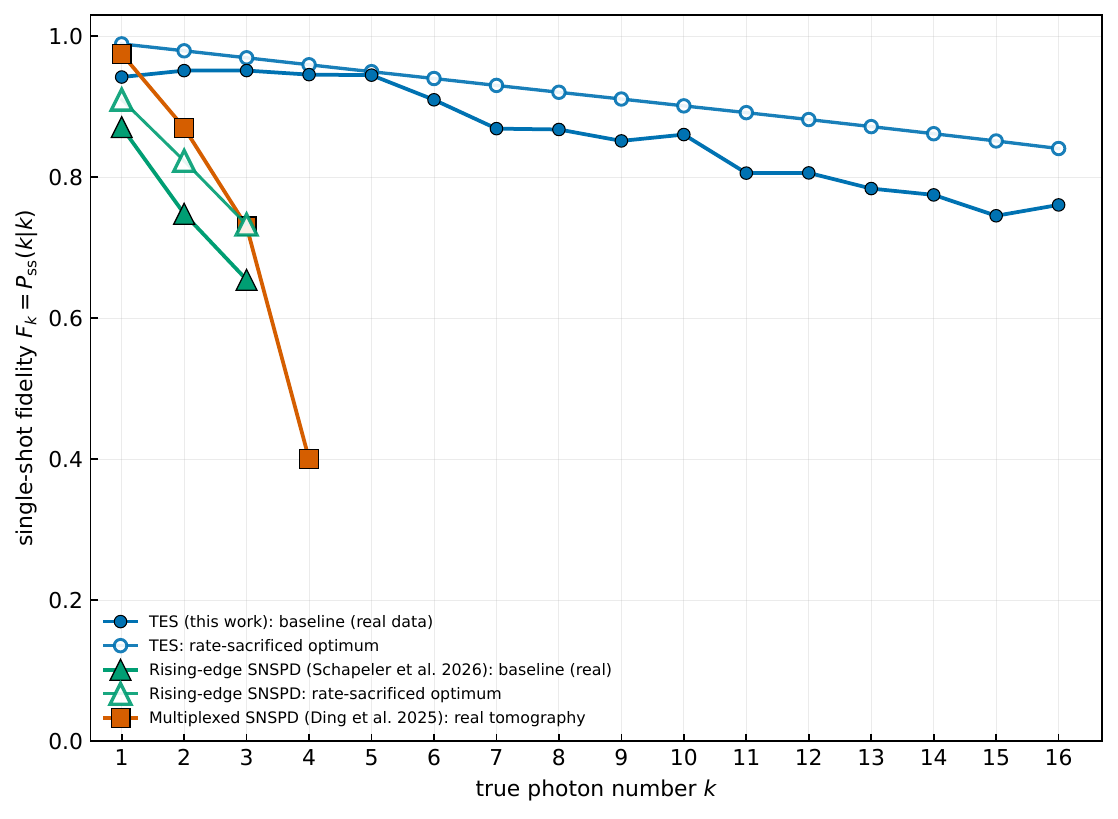}
  \caption{Comparison of the improvement obtainable by adjusting the accept region, for the TES and the rising-edge-timing SNSPD, across the heralded photon number $k=1$ to $16$. The TES (blue) shows the baseline based on real calibration data (solid circles, solid line) and the level reached once the accept region is narrowed to the point where further narrowing no longer improves fidelity (open circles, solid line), across the full range $k=1$--$16$. This reached level is a calculated value based on real calibration data and real detection efficiency, not a value obtained by re-measuring with the accept region actually narrowed. The rising-edge-timing SNSPD (green, Schapeler \emph{et al.}~\cite{Schapeler2026}) shows the baseline given by the diagonal of the measured POVM (solid triangles, $k=1,2,3$) and the level reached by combining that paper's yield-sacrificed optimized result with the real detection efficiency ($\eta=91\,\%$) (open triangles). This detector saturates around $k\approx6$--$7$, beyond which it has not been characterized. The multiplexed SNSPD (orange, Ding \emph{et al.}~\cite{Ding2025}) shows measured tomography values ($k=1,2,3,4$); because its misidentification arises only in one direction, it has no corresponding reached level.}
  \label{fig:detector_comparison}
\end{figure}

\begin{sloppypar}
Evaluated on this common framework and common metric $F_k$, at the single-photon point $k=1$ the multiplexed SNSPD (measured value $97.5\,\%$) slightly exceeds our TES (measured value $94.2\,\%$), with the rising-edge-timing SNSPD (measured value $87.1\,\%$) close behind. Across the broader low-photon-number range relevant to DV architectures, however, the relative performance depends on the required photon number, with the measured multiplexed-SNSPD fidelity degrading substantially toward $k=4$ (Fig.~\ref{fig:detector_comparison}). At the CV-relevant $k=16$, by contrast, the multiplexed SNSPD is fundamentally unable to reach this regime because of the theoretical ceiling set by detection efficiency alone ($\eta^{16}\approx72.4\,\%$; the actual calculated value is $0.75\,\%$), while the rising-edge-timing SNSPD simply lies outside its characterized range. The TES maintains $76.1\,\%$ fidelity even in this regime.
\end{sloppypar}

Figure~\ref{fig:detector_comparison} overlays, for the two detectors where improvement from adjusting the accept region can actually be confirmed -- the TES and the rising-edge-timing SNSPD -- the baseline (no rejection, $\tau=1$) and the level reached by sacrificing yield, across $k=1$ to $16$ wherever data exist. For the TES, we show the baseline $F_k$ (solid circles, solid line) and the level reached once the accept region is narrowed to the point where further narrowing no longer improves fidelity (open circles, solid line). This reached level is a calculated value based on real calibration data ($\mu_n,\sigma_n$) and the real detection efficiency ($\eta=99\,\%$); it is close to, but not exactly equal to, the theoretical value $\eta^k$ ($84.1\,\%$ calculated versus $\eta^{16}=85.1\,\%$ at $k=16$), because even maximal narrowing of the accept region does not drive the resolution-driven confidence all the way to unity (it saturates at $98.7\,\%$, for example, at $k=16$). For the rising-edge-timing SNSPD, we similarly show the baseline Schapeler \emph{et al.} report (solid triangles) and the level reached by combining their yield-sacrificed optimized result with that paper's detection efficiency ($\eta=91\,\%$) (open triangles), for $k=1,2,3$. For both detectors, the improvement from the solid to the open markers is an independent confirmation, across architecturally distinct detectors, of the same underlying physical mechanism: adjustment of the accept region. Because misidentification in the multiplexed SNSPD (Ding \emph{et al.}~\cite{Ding2025}) arises only in one direction, its measured tomography values ($k=1,2,3,4$, open squares) have no corresponding reached level.

This result demonstrates concretely the practical value of the present framework: even though DV and CV architectures make different demands, both can be compared consistently within a common framework. At the same time, it shows directly that no single detector technology is universally best -- the appropriate technology depends on the required photon-number range -- and that whether room for improvement (via adjustment of the accept region) exists at all likewise depends on the detector's operating principle.

\section*{Conclusion}

This work identifies a gap in how PNR-detector performance has been evaluated: single-shot misidentification due to finite energy resolution is, in principle, not explicitly separated from efficiency loss in conventional (ensemble-averaged) definitions of fidelity. We built an evaluation framework that corrects this gap. Incorporating a resolution confusion matrix into the measured POVM yields a single-shot POVM, whose diagonal element $F_k=P_{\mathrm{ss}}(k|k)$ we defined as a single-shot fidelity that requires no separate reference theoretical model. Under this definition, we showed that detection-efficiency loss and resolution-driven misidentification are asymmetric in a specific sense: whether or not a judgment can be held back and converted into an erasure. Detection efficiency imposes an unrecoverable, ``hard'' performance limit, whereas resolution imposes a ``soft'' performance limit that can be improved, at the cost of generation rate, by adjusting the accept region.

Applying this framework to real data from three architecturally distinct detectors -- a multiplexed SNSPD, a rising-edge-timing SNSPD, and a TES -- we found that which detector is preferable switches sharply between the low-photon-number regime demanded by DV architectures and the high-photon-number regime demanded by CV architectures. The inability of the multiplexed SNSPD to reach the high-photon-number regime traces to an unrecoverable theoretical ceiling set by detection efficiency alone, whereas the rising-edge-timing SNSPD's inability to do so, while currently limited by peak overlap, is amenable to quantitative improvement through adjustment of the accept region, as we demonstrated using real data. Detectors of different architectures can only be meaningfully compared, and fundamental limits distinguished from improvable ones, on a common metric such as single-shot fidelity.

This framework presupposes neither a particular detector's operating principle nor a particular application: it applies wherever a detector's output retains any gradient of confidence about the true photon number. We plan to extend its scope of applicability further, both by applying it to other CV applications (such as verification via Gaussian boson sampling) and by incorporating single-shot fidelity into fidelity simulations of GKP-breeding states.

\bibliographystyle{unsrt}
\bibliography{refs}

\end{document}